\documentclass[a4paper,12pt]{article}
\usepackage{tikz}
\usetikzlibrary{arrows,snakes,backgrounds}
\usepackage{jheppub} % for details on the use of the package, please
\usepackage{amsmath}
\usepackage{graphicx}
\usepackage{hyperref}
\usepackage{lmodern}
\usepackage{amssymb}
\usepackage{booktabs}

\usepackage{amsfonts}
\usepackage{amssymb}
\usepackage{amsthm}
\usepackage{tikz}

\newcommand{\be}{\begin{equation}}
\newcommand{\eeq}{\end{equation}}
\newcommand{\bea}{\begin{eqnarray}}
\newcommand{\eea}{\end{eqnarray}}
\newcommand{\ba}{\begin{array}}
    \newcommand{\ea}{\end{array}}
\def\nn{\nonumber}

\newcommand{\ee}{\end{equation} }

\newcommand{\one}{{\rm 1\kern -.9mm l}}

\def\CD {{\cal D}}

\def\CF {{\cal F}}

\def\CM {{\cal M}}

\def\CO {{\cal O}}

\title{Quantum-Mechanical Interpretation of Riemann Zeta Function Zeros}

\author{ George Savvidy and Konstantin Savvidy}

\affiliation{Institute of Nuclear and Particle Physics\\
 Demokritos National Research Centre,  Athens, Greece}
\emailAdd{ savvidy@inp.demokritos.gr}

\abstract{ We demonstrate that the Riemann zeta function zeros define the position and the widths of the resonances of the quantised Artin dynamical system.
The Artin dynamical system is defined on the fundamental region of the modular group on the Lobachevsky plane. It has a finite volume and an infinite extension in the vertical direction that correspond to a cusp.  In classical regime the geodesic flow in the fundamental region represents one of the most chaotic dynamical systems, has mixing of all orders, Lebesgue spectrum and non-zero Kolmogorov entropy.  In quantum-mechanical regime the system can be associated with the narrow infinitely long waveguide stretched out to infinity along the vertical axis  and a cavity resonator attached to it at the bottom.  That suggests  a physical interpretation of the Maass automorphic wave function in the form of an incoming plane wave of a given energy entering the resonator,  bouncing inside the resonator and scattering to infinity.  As  the energy of the incoming wave comes close to the eigenmodes of the cavity a pronounced resonance behaviour shows up in the scattering amplitude.}

\preprint{NRCPS-HE-66-2018 }

\notoc

\begin{document}
    \maketitle
    \flushbottom

\section{\it Introduction}

\begin{figure}
\hspace*{-1cm} 
\begin{tikzpicture}[scale=2]
\clip (-4.3,-0.4) rectangle (4.5,2.8);
\draw[step=.5cm,style=help lines,dotted] (-3.4,-1.4) grid (3.4,2.6);
\draw[,->] (-2.3,0) -- (2.4,0); \draw[->] (0,0) -- (0,2.3);
\foreach \x in {-1,-0.5,0,0.5,1}
\draw (\x cm,1pt) -- (\x cm,-1pt) node[anchor=north] {$\x$};
\foreach \y in {1}
\draw (1pt,\y cm) -- (-1pt,\y cm) node[anchor=east] {$\y$};
\draw (0,0) arc (0:180:1cm);
\draw (1,0) arc (0:180:1cm);
\draw (2,0) arc (0:180:1cm);
\draw (0.5,0)--(0.5,0.86602540378);
\draw (-0.5,0)--(-0.5,0.86602540378);
\draw[ ultra thick, blue] (0.5,0.86602540378)--(0.5,2.1);
\draw[ ultra thick, blue] (-0.5,0.86602540378)--(-0.5,2.1);
\draw[ultra thick, blue] (0.5,0.86602540378) arc (60:120:1cm);
\draw (-0.25,1.8) node{ $\CF$};
\draw (0,2.4) node{ $D$};
\draw (0,2.4) node{ $D$};
\draw (-0.6,0.7) node{ $A$};
\draw (0.4,0.7) node{ $B$};
\draw (-0.1,0.85) node{ $C$};
\draw[fill] (-0.5,0.86602540378) circle [radius=0.02];
\draw[fill] (0,1) circle [radius=0.02];
\draw[fill] (0.5,0.86602540378) circle [radius=0.02];
\draw[thin,teal] (-0.5,1.1)  arc (60:30:0.41cm);
\draw[thin,teal] (0.35,0.952)  arc (150:120:0.41cm);
\draw [thin,teal] (-0.43,0.98) node{ $\alpha$};
\draw [thin,teal] (0.43,0.98) node{ $\beta$};
\end{tikzpicture}
\caption{The  non-compact fundamental region $\CF$ of a finite area $\pi/3$ is represented by the hyperbolic triangle $ABD$.  The vertex $D$ is at infinity of the $y$ axis and corresponds to a  cusp. The edges of the triangle are the arc $AB$,  the rays $AD$   and $BD$. The points on the edges $AD$ and $BD$ and the points of the arks $AC$ with $CB$ 
are identified by the transformations $w=z+1$ and $w = -1/z$ in order to form a {\it closed  non-compact surface $\bar{\CF}$ of sphere topology}
by  "gluing together" the opposite edges of the modular triangle \cite{Babujian:2018xoy}.}
\label{fig1} 
\end{figure}
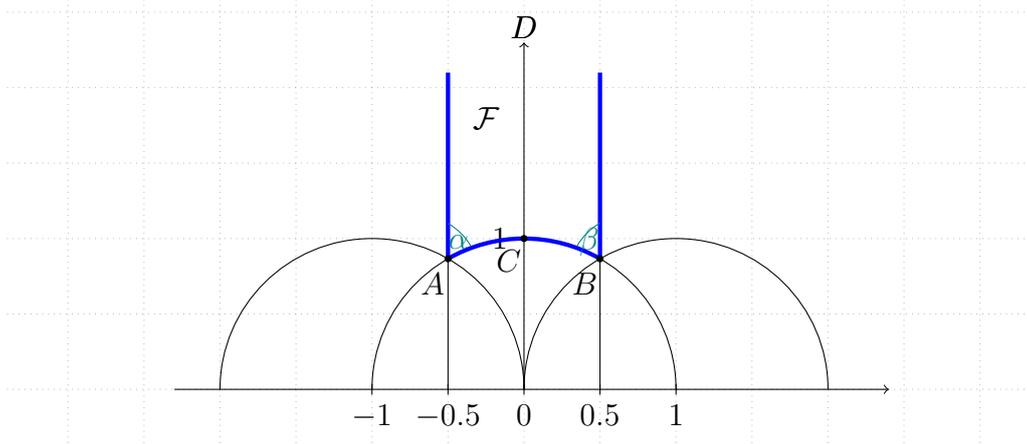

Hyperbolic systems  have exponential instability of their  trajectories and  as such  represent the most natural  chaotic dynamical systems \cite{anosov}. Of special interest are systems which are defined on closed surfaces  of  the Lobachevsky plane of constant negative curvature.  An example of such system was introduced in 1924 by the mathematician Emil Artin \cite{Artin}.  The  dynamical system is defined on the fundamental region of the Lobachevsky plane which is obtained by the identification of points  congruent with respect to the modular group $\Gamma=SL(2,Z)$,  a  discrete subgroup of the Lobachevsky plane isometries \cite{Poincare,Poincare1,Fuchs}.  The fundamental region $\CF$ in this case is {\it an infinitely long  non-compact hyperbolic triangle of finite area} shown in Fig.\ref{fig1}. The geodesic trajectories are bounded to propagate on the fundamental region and represent one of the most chaotic dynamical systems with exponential instability  of its trajectories, mixing of all orders, Lebesgue spectrum and non-zero Kolmogorov entropy 
\cite{hadamard,hedlund,anosov,bowen0,kolmo,kolmo1,sinai3,ruelle,hopf,Hopf,Gelfand,Poghosyan:2018efd,yangmillsmech,Savvidy:1982jk} .  

\begin{figure}
 \centering
 \includegraphics[angle=0,width=7cm]{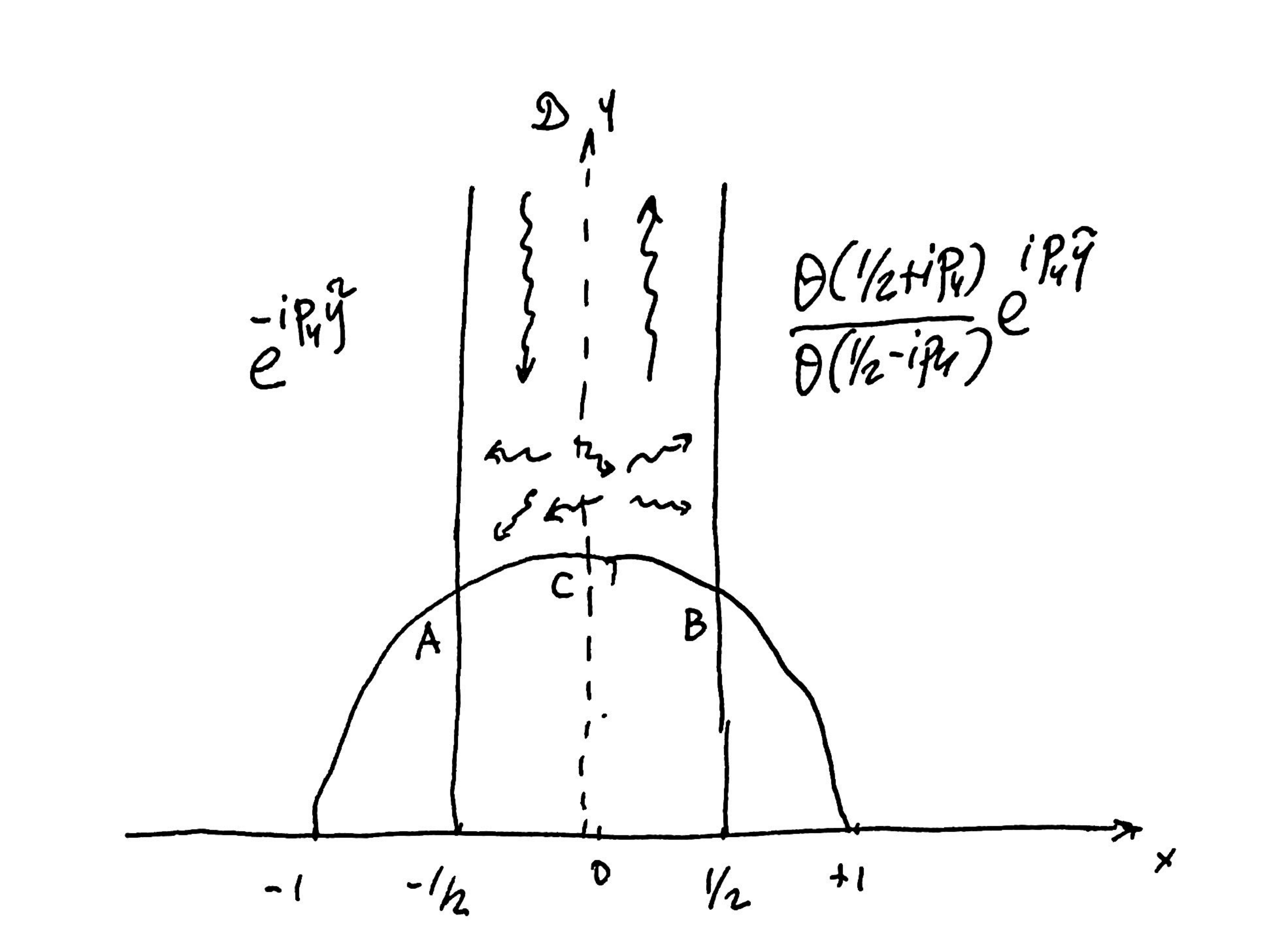}
\caption{ The incoming and outgoing plane waves. The plane wave  $e^{-i p_y \tilde{y}  }$ incoming from infinity of the $y$ axis on Fig. \ref{fig1}  ( the vertex $\CD$)  elastically scatters on the boundary $ACB$ of the fundamental triangle  $\CF$ on Fig. \ref{fig1}.  The reflection amplitude is a pure phase and is given by the expression in front of the outgoing plane wave $e^{  i p_y \tilde{y}}$.
The rest of the wave function describes the standing waves in the $x$ direction between the boundaries $x=\pm 1/2$ with the amplitudes, which are exponentially decreasing.}
\label{fig2}
\end{figure}

\begin{figure}
 \centering
 \includegraphics[angle=0,width=6cm]{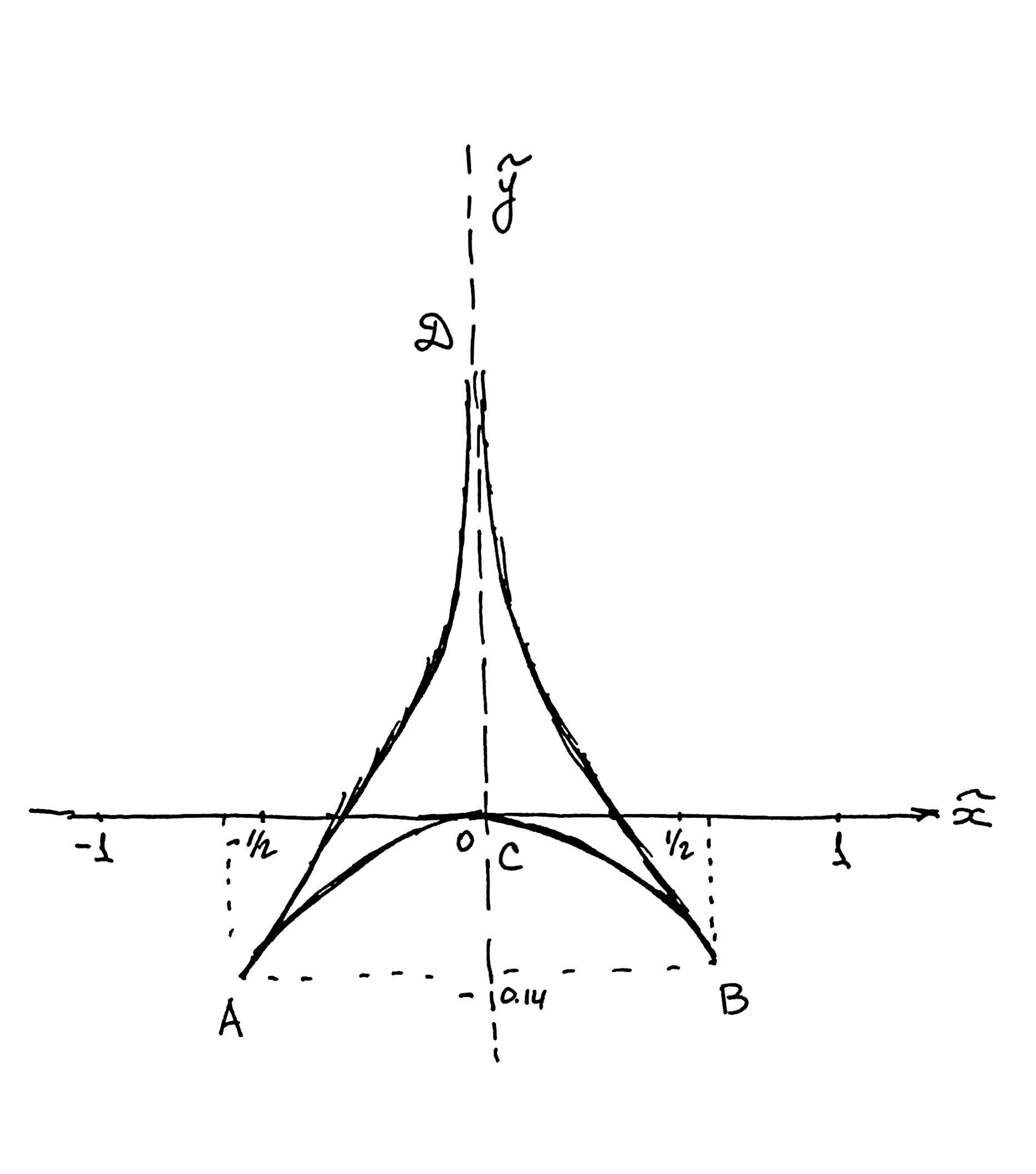}
\caption{ The system can be described as a narrow (\ref{111}) infinitely long waveguide stretched to infinity along the vertical dierection  and a cavity resonator attached to it at the bottom $ACB$.}
\label{fig3}
\end{figure}

In a recent article \cite{Babujian:2018xoy} the authors investigated the behaviour of  
quantum-mechanical correlation functions of the quantised Artin system \cite{Faddeev:1969su}. 
The solution of the time-independent Schr\"odinger equation 
$
H\psi = E \psi
$
\bea
-y^2(\partial_x^2+\partial_y^2)\psi= E \psi
\eea 
with periodic boundary conditions on the wave function with respect to the modular group 
in the fundamental region $\bar{\CF}$ shown in Fig. \ref{fig1}    is:
\bea\label{invariance}
\psi\Big(\frac{a z+b}{c z+d}\Big)=\psi(z),~~~\left(
\begin{array}{cc}
a&b\\c&d
\end{array}
\right) \in SL(2,Z), 
\eea
and it is defined by the Maass non-analytical automorphic function \cite{maass,roeleke,selberg1,selberg2,Faddeev,Faddeev1,hejhal2,winkler,hejhal,hejhal1}.
One important 
novelty  introduced in the article \cite{Babujian:2018xoy} was the representation of the Maass wave function \cite{maass} in terms of the  natural physical variable $\tilde{y}$ which represents the distance in the vertical direction of the Lobachevsky plane  $\int dy/y =  \ln y = \tilde{y} $   and of the corresponding momentum $p$ \cite{Babujian:2018xoy}. This 
allows to represent the energy eigenfunctions obtained by Maass  in the form which is appealing to the physical intuition  \cite{Babujian:2018xoy}:
\bea\label{alterwave00}
\psi_{p} (x,\tilde{y})  
= e^{-i p \tilde{y}  }+{\theta(\frac{1}{2} +i p) \over \theta(\frac{1}{2} -i p)}   \, e^{  i p \tilde{y}} + { 4  \over  \theta(\frac{1}{2} -i p)}   \sum_{l=1}^{\infty}\tau_{i p}(l)
K_{i p }(2 \pi  l e^{\tilde{y}} )\cos(2\pi l x), \nn\\
\eea
where 
\bea\label{thata1}
\theta(s)=\pi ^{-s} \zeta (2 s) \Gamma (s)
\eea  
is a product of Riemann zeta  and  gamma functions,  $K$ is  the modified Bessel's 
 function  
\be\label{modifiedB}
K_{i p}(y) = {1\over 2} \int^{\infty}_{-\infty} e^{- y \cosh t} e^{i p t}   dt ,
\ee
and 
\bea
\tau_{i p}(n )=\sum_{a \cdot b=n}\left(\frac{a}{b}\right)^{ip} .
\eea
The first two terms of the wave function (\ref{alterwave00}) describe the incoming and outgoing plane waves. The plane wave  $e^{-i p \tilde{y}  }$ incoming from infinity of the $y$ axis on Fig. \ref{fig2},  the vertex $D$,   elastically scatters on the boundary $ACB$ of the fundamental triangle  $\CF$.  The reflection amplitude is a pure phase and is given by the expression in front of the outgoing plane wave $e^{  i p \tilde{y}}$ :
\be\label{Smatrix}
S={\theta(\frac{1}{2} +i p) \over \theta(\frac{1}{2} -i p)} = \exp{[i\, 2\, \delta(p)]}.
\ee
The rest of the wave function describes the standing waves $\cos(2\pi l x)$  in the $x$ direction between the boundaries $x=\pm 1/2$
with the amplitudes $K_{i p }(2 \pi  l e^{\tilde{y} })$, which are exponentially decreasing with index $l$. The continuous energy spectrum is given by the formula   \cite{Babujian:2018xoy}
\be\label{energy}
E=   p^2  + \frac{1}{4} .
\ee
{\it In physical terms the system can be described as a narrow infinitely long waveguide stretched out to infinity along the vertical dierection  and a cavity resonator attached to it at the bottom  $ACB$ } 
(see Fig.\ref{fig3}).  In order to support this interpretation we shall calculate the area of the fundamental region which is below the fixed coordinate $y_0= e^{\tilde{y}_0}$:
\be
\text{Area}(\CF_0)=\int _{-\frac{1}{2}}^{\frac{1}{2}} dx
\int _{\sqrt{1-x^2}}^{y_0 }\frac{dy}{y^2}  =\frac{\pi }{3}- e^{-\tilde{y}_0}\, ,
\ee 
and confirm that the area above  the ordinate $\tilde{y}_0$ is exponentially small: $e^{-\tilde{y}_0}$. The horizontal ( $dy =0$) size of the fundamental region also decreases exponentially in the vertical direction:
\be\label{111}
L_0=\int  ds =\int \frac{\sqrt{dx^2+dy^2}}{ y}=  \int _{-\frac{1}{2}}^{\frac{1}{2}} {dx\over y_0 } = e^{-\tilde{y}_0}.
\ee 
One can suggest therefore the following physical interpretation of the Maass wave function (\ref{alterwave00}): The incoming plane wave $e^{- i p \tilde{y}}$ of energy $E=p^2  + \frac{1}{4} $ enters the "cavity resonator", 
bouncing  back into the outgoing plane wave at infinity $e^{i p \tilde{y}}$.  As the energy of the incoming wave $E =p^2  + \frac{1}{4}$ becomes close to the eigenmodes of the cavity  resonator one should expect a pronounced resonance behaviour of the scattering amplitude. 

To trace such behaviour let us 
consider the analytical continuation of the Maass wave function (\ref{alterwave00}) to the complex  energies $E$.
The analytical continuation of the scattering 
amplitudes as a function of the energy $E$ considered as a complex variable allows to establish important spectral properties of the quantum-mechanical system. In particular,  the method of analytic continuation allows to determine the real and complex S-matrix poles. The real  poles on the physical sheet correspond  to the  discrete energy levels and the complex poles on the second sheet below the cut correspond  to the resonances in the quantum-mechanical system  \cite{landauqmech} .  

Indeed, the asymptotic form of the wave function  can be represented in the following form: 
\be\label{reswave4}
\psi = A(E)\, e^{i p \tilde{y}} + B(E)\, e^{-i p \tilde{y}} , ~~~p = \sqrt{E-1/4} ,
\ee
and  to make the functions $A(E) $ and $B(E) $ single-valued one should cut the complex plane 
along the real axis \cite{landauqmech} starting from $E=1/4$. The complex plane with a cut so defined is called a physical sheet.  To left from the cut,  at  energies  $E_0 < 1/4$,  the wave function will take the following form: 
\be\label{reswave1}
\psi = A(E)\, e^{- \sqrt{\vert E-1/4 \vert } \tilde{y}} + B(E)\, e^{ \sqrt{\vert E-1/4 \vert} \tilde{y}}, 
\ee
where the exponential factors are real and one of them decreases and the other one increases at $\tilde{y} \rightarrow \infty$.  The  bound states are characterised  by the fact that the corresponding wave function tends to zero at infinity $\tilde{y} \rightarrow \infty$.  
This means that the second term in (\ref{reswave1})   should be absent, and a discrete energy level $E_0 < 1/4$ corresponds to a zero of the $B(E)$ function \cite{landauqmech}: 
\be\label{zeros2}
B(E_0) =0.~~~
\ee
Because the energy eigenvalues are real, all zeros of $B(E)$ on the physical sheet are real.

 Now consider a system which is unstable and therefore does not have a pure discrete spectrum, the motion of the system is unbounded  and the energy spectrum is continuous \cite{landauqmech}. The energy spectrum is quasi-discrete, consisting of smeared levels of  a width $\Gamma$.  In describing  such states one should consider the wave functions which are diverging at infinity,  describing a wave packet moving to infinity.  Thus the boundary condition at infinity requires the presence of only outgoing waves. This boundary condition involves complex quantities and the energy eigenvalues in general are also complex \cite{landauqmech}. With such boundary conditions the Hermitian energy operators can have complex eigenvalues of the form \cite{landauqmech}
\be\label{compeigen1}
E = E_0 - i {\Gamma \over 2},
\ee
where $E_0$ and $\Gamma$ are both real and positive.  
\begin{figure}
 \centering
 \includegraphics[angle=0,width=8cm]{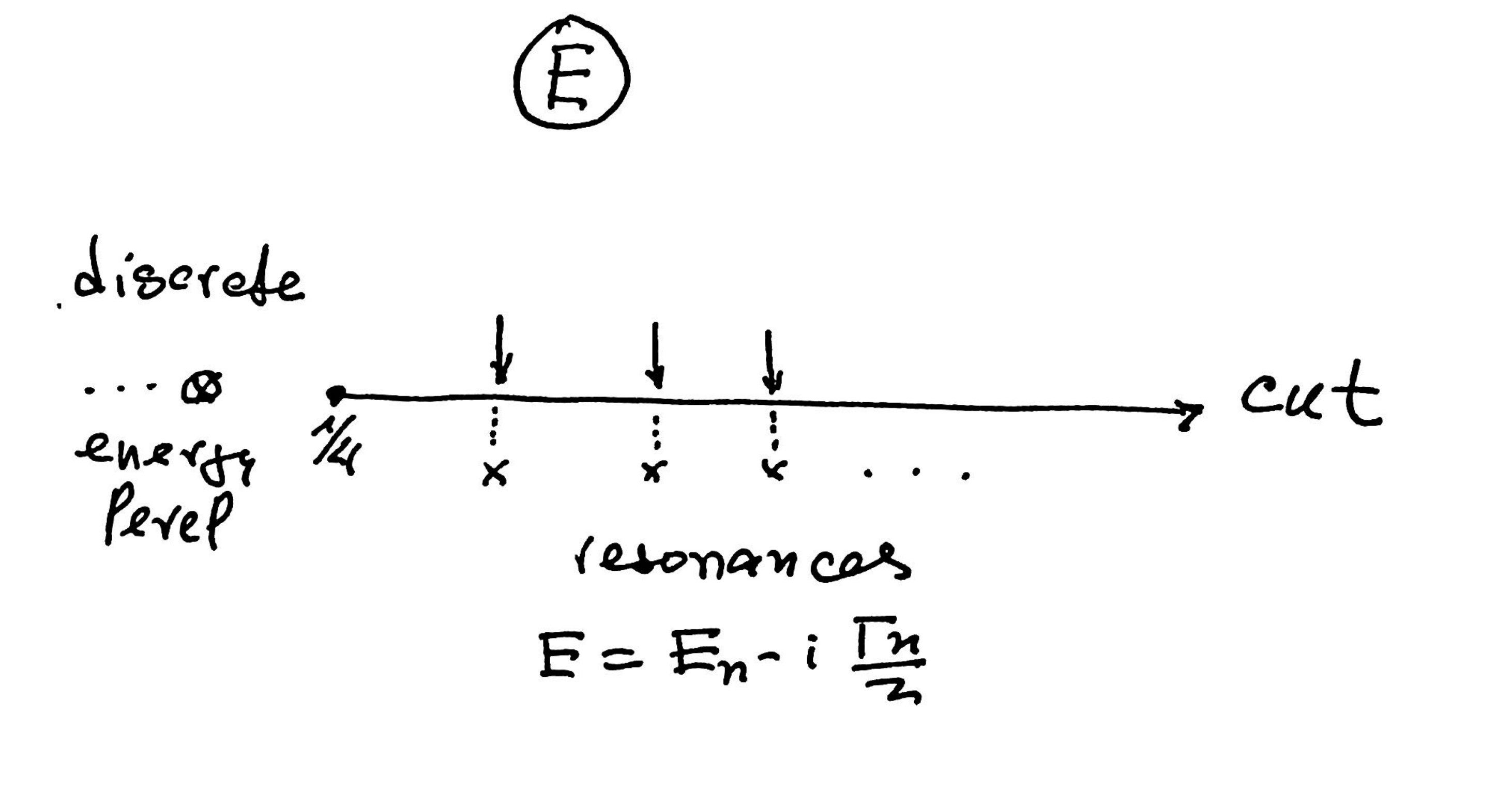}
\caption{ The resonances  $E_n - i {\Gamma_n \over 2}$ are located under the right hand side of the real axis.}
\label{fig4}
\end{figure}
The condition which defines the complex energy eigenvalues (\ref{compeigen1}) lies in the requirement  that at $E = E_0 - i {\Gamma \over 2}$ the incoming wave $ e^{-i p \tilde{y}} $ in (\ref{reswave4})   is absent \cite{landauqmech}:     
\be\label{zerosquasi1}
B(E_0 - i {\Gamma \over 2}) =0.
\ee
The point $E_0 - i {\Gamma \over 2}$ is located under the right hand side of the real axis, see 
Fig.\ref{fig4}. In order to reach that point without leaving the physical sheet one should move from the upper side of the cut  anticlockwise.  However in that case, the phase of the wave function changes its  sign and the outgoing wave transforms into the incoming wave. In order to keep the outgoing character of the wave function one should cross  the cut strait into the  second sheet Fig.\ref{fig4}.  Expanding the function 
$B(E)$ near the quasi-discrete energy level (\ref{compeigen1}) as $B(E) =(E - E_0 + {i \Gamma \over 2}) b +...$ one can get 
\be
\psi~~ \approx~~ b^* (E - E_0 - {i \Gamma \over 2})  e^{i p \tilde{y}} +b  (E - E_0 + {i  \Gamma \over 2})  e^{-i p \tilde{y}}
\ee
and the S-matrix  will take the following form \cite{landauqmech}
\be\label{resonphase}
S=e^{2 i \delta} = {E - E_0 - i \Gamma / 2 \over E - E_0 +i \Gamma / 2} e^{2 i \delta_0},
\ee
where $e^{2 i \delta_0} =  b^* / b $. One can observe that moving throughout the resonance region the phase is changing by $\pi$.

Let us now consider the asymptotic  behaviour of the wave function (\ref{alterwave00}) at large $\tilde{y}$.  The conditions (\ref{zeros2}) and  (\ref{zerosquasi1}) of the absence of incoming wave takes the form:
\be
 \theta(\frac{1}{2} -i p) =0
\ee
and due to  (\ref{thata1}):
\bea 
\theta(\frac{1}{2} -i p)=  {\zeta (1- 2 i p) \Gamma (\frac{1}{2} -i p) \over  \pi ^{\frac{1}{2} -i p} }=0.
\eea  
The solution of this equation can be expressed in terms of zeros  of the Riemann zeta function
\cite{Riemann}:
\be
\zeta(\frac{1}{2} - i u_n) =0,~~~~  n=1,2,....~~~~u_n > 0.
\ee
Thus one should solve the equation 
\be
1- 2 i p_n = \frac{1}{2} - i u_n~.
\ee 
The location of poles is therefore at the following values of the complex momenta 
\be\label{poles}
p_n = {u_n \over 2} - i \,{1 \over 4} \, ,  ~~~~~  n=1,2,..... 
\ee
and at the corresponding complex energies (\ref{energy}) :
\be\label{resonances34} 
E = p^2_n +{1 \over 4}~ =~ ({u_n \over 2} -   \,{1 \over 4}\,i)^2  +{1 \over 4} ~= ~{u^2_n \over 4} + {3\over 16}
- i \, {u_n \over 4}.
\ee
Thus one can observe that there are resonances (\ref{compeigen1})
\be
E = E_n - i {\Gamma_n \over 2}
\ee
at the following energies and of the corresponding widths (\ref{resonances34}):
\be\label{exactresonance}
E_n = {u^2_n \over 4} + {3\over 16},~~~~~~\Gamma_n = {u_n \over 2} . 
\ee
The ratio of the width to the energy tends to zero  \cite{Riemann}:
\be
{ \Gamma_n  \over E_n} = {u_n \over 2}/ ({u^2_n \over 4} + {3\over 16}) \approx {2 \over u_n} 
\rightarrow ~~0
\ee
and the resonances become infinitely narrow. The ratio of the width to the energy spacing between nearest levels is  
\be
{\Gamma_n \over  E_{n+1} - E_n } = {2 u_n \over (u_{n+1} + u_n)(u_{n+1} - u_n)} \approx 
{1 \over u_{n+1} - u_n}  .
\ee
As far as the zeros of the zeta function have the property to  "repel", the difference $u_{n+1} - u_n$ can vanish with small probability \cite{Turing,Gourdon}.

Thus one can conjecture the following representation of the S-matrix (\ref{Smatrix}):
\be\label{Smatirxphase}
S=e^{2 i\, \delta} =  {\theta(\frac{1}{2} +i p) \over \theta(\frac{1}{2} -i p)} = \sum^{\infty}_{n=1}{E - E_n - i \Gamma_n / 2 \over E - E_n +i \Gamma_n / 2}~ e^{2 i \delta_n}
\ee
with yet unknown phases $\delta_n$. In order to justify the above representation of the S-matrix we shall try to find the location of the poles on the second Riemann sheet by using expansion of the S-matrix
(\ref{Smatrix}) at the "bumps" which occur  along the real axis at energies  
\be \label{approxresonance}
E_n  = {u^2_n \over 4}  +{3\over 16}~.
\ee
The expantion will take the following form:
\bea\label{Smatirxphase1}
S\vert_{E \approx E_n} &  
= &  {\theta(\frac{1}{2} +i \sqrt{E - {1\over 4}} ) \over 
\theta( \frac{1}{2} -i  \sqrt{E - {1\over 4} }) } \vert_{E \approx E_n}~\nn\\
&=&~{ \theta(\frac{1}{2} +i \sqrt{E_n - {1\over 4}} )  +   \theta^{'}(\frac{1}{2} 
+i \sqrt{E_n - {1\over 4}} ) ~ (E - E_n)
 \over  \theta(\frac{1}{2} -i \sqrt{E_n - {1\over 4}} )  +   \theta^{'}(\frac{1}{2} -i \sqrt{E_n - {1\over 4}} ) ~  (E - E_n)} \nn\\
&=& 
 {    E - E_n + \theta(\frac{1}{2} +i \sqrt{E_n - {1\over 4}} ) /\theta^{'}(\frac{1}{2} 
 +i \sqrt{E_n - {1\over 4}} ) 
 \over    E - E_n + \theta(\frac{1}{2} -i \sqrt{E_n - {1\over 4}} )/\theta^{'}(\frac{1}{2} 
 -i \sqrt{E_n - {1\over 4}} ) } ~~ {\theta^{'}(\frac{1}{2} +i \sqrt{E_n - {1\over 4}} )   \over \theta^{'}(\frac{1}{2} -i \sqrt{E_n - {1\over 4}} )   }  \nn\\ 
 &\equiv &  ~~~
 {    E - E^{'}_n - i \Gamma^{'}_n/2   
 \over    E - E^{'}_n + i \Gamma^{'}_n/2  }~~  e^{2 i \delta^{'}_n}~,\nn\\
 \eea
where
\bea\label{approxresonance1}
E^{'}_n - i \Gamma^{'}_n/2 = E_n - {\theta(\frac{1}{2} -i \sqrt{E_n - {1\over 4}} ) 
\over \theta^{'}(\frac{1}{2} -i \sqrt{E_n - {1\over 4}} ) } ,~~~~ e^{2 i \delta^{'}_n}=
 {\theta^{'}(\frac{1}{2} +i \sqrt{E_n - {1\over 4}} )   \over \theta^{'}(\frac{1}{2} 
 -i \sqrt{E_n - {1\over 4}} )   } , 
\eea 
 thus
\bea
E^{'}_n - E_n = - \Re {\theta(\frac{1}{2} -i \sqrt{E_n - {1\over 4}} ) 
\over \theta^{'}(\frac{1}{2} -i \sqrt{E_n - {1\over 4}} ) } ~~~- i  \Gamma^{'}_n/2 = - \Im  {\theta(\frac{1}{2} -i \sqrt{E_n - {1\over 4}} ) 
\over \theta^{'}(\frac{1}{2} -i \sqrt{E_n - {1\over 4}} ) } 
\eea
and all quantities $E^{'}_n$ , $ \Gamma^{'}_n/2$ and $\delta^{'}_n$ are real.  

Let us consider the first ten zeros of the zeta function which are known numerically \cite{Turing,Gourdon}.  Using  that information one can calculate the position of the resonances and their widths (\ref{exactresonance}).  In the Table \ref{table1} we present the values of energies and the widths of the resonances given by exact formula 
(\ref{exactresonance}). In  Table \ref{table2}  the energies and widths are calculated using the approximation formulas   (\ref{approxresonance1}). As one can see the approximation is consistent with the exact values within the two precent deviation.
\vspace{4cm}

\begin{table} 
  \centering
 \begin{tabular}{@{}ccccc @{}} 
     %\begin{tabular}{ccccc}  
     \toprule
      Position of zeros &~ Energies $E_n$ & Width $\Gamma_n$   \\
        \midrule
   $u_1$ = 14.1347&~~~~~ $E_1$ = 50.1351&~~~~~$\Gamma_1$ = 3.53368    \\
       $u_2$ = 21.0220&~~~~~ $E_2$ = 110.669 &~~~~~$\Gamma_2$ =  5.25551\\ 
     $u_3$ = 25,0109&~~~~~ $E_3$ = 156,573&~~~~~$\Gamma_3$ =  6.25271     \\      
       $u_4$ =  30,4249&~~~~~ $E_4$ = 231,606&~~~~~$\Gamma_4$ =  7,60622     \\      
      $u_5$ = 32,9351&~~~~~ $E_5$ = 271,367&~~~~~$\Gamma_5$ =  8.23377  \\      
    $u_6$ = 37,5862&~~~~~ $E_6$ = 353,368&~~~~~$\Gamma_6$ =  9.39654     \\       
$u_7$ =  40,9187&~~~~~ $E_7$ = 418,773&~~~~~$\Gamma_7$ =  10.2297    \\ 
$u_8$ =  43,3271&~~~~~ $E_8$ = 469,496&~~~~~$\Gamma_8$ =  10.8318    \\ 
$u_9$ =  48,0052&~~~~~ $E_9$ = 576,311&~~~~~$\Gamma_9$ =  12.0013   \\
$u_{10}$ =  49,7738&~~~~~ $E_{10}$ = 619,546&~~~~~$\Gamma_{10}$ =  12.4435   \\
.................&............&~~~.\\
 \bottomrule

      \end{tabular}
        \caption{The table presents the numerical values $u_n$ of the zeros of the Riemann zeta function. The corresponding energies and widths (\ref{exactresonance}) are given on the second and the third columns. }
      \label{table1}
\end{table}
\begin{table} 
  \centering
 \begin{tabular}{@{}ccccc @{}} 
     %\begin{tabular}{ccccc}  
     \toprule
      Position of zeros &~ Energies $E^{'}_n$ & Width $\Gamma^{'}_n$    \\
        \midrule
$u_1$ = 14.1347&~~~~~ $E^{'}_1$ = 51.2732 &~~~~~$\Gamma^{'}_1$ =   3.05908\\
$u_2$ = 21,0220&~~~~~ $E^{'}_2$ = 112.487&~~~~~$\Gamma^{'}_2$ =   4.32077 \\ 
$u_3$ = 25,0109&~~~~~ $E^{'}_3$ = 158.363 &~~~~~$\Gamma^{'}_3$ =  5.42025    \\      
 $u_4$ =  30,4249&~~~~~ $E^{'}_4$ =  234.382 &~~~~~$\Gamma^{'}_4$ =  5.79733 \\        
$u_5$ = 32,9351&~~~~~ $E^{'}_5$ = 273.225 &~~~~~$\Gamma^{'}_5$ =  7.20321 \\       
 $u_6$ = 37,5862&~~~~~ $E^{'}_6$ = 356.546 &~~~~~$\Gamma^{'}_6$ = 7.5043 \\        
$u_7$ =  40,9187&~~~~~ $E^{'}_7$ = 422.097  &~~~~~$\Gamma^{'}_7$ =  7.99925  \\ 
$u_8$ =  43,3271&~~~~~ $E^{'}_8$ = 471.764 &~~~~~$\Gamma^{'}_8$ =  9.44046  \\ 
$u_9$ =  48,0052&~~~~~ $E^{'}_9$ = 580.782 &~~~~~$\Gamma^{'}_9$ =  8.29622  \\
$u_{10}$ =  49,7738&~~~~~ $E^{'}_{10}$ = 621.9 &~~~~~$\Gamma^{'}_{10}$ =  10.4703    \\
.................&............& ....\\
 \bottomrule

      \end{tabular}
        \caption{The table presents the numerical values $u_n$ of the zeros of the Riemann zeta function.  The energies and widths are calculated by using the approximation formulas   (\ref{approxresonance1}). }
      \label{table2}
\end{table}

{\bf Acknowledgments.}
We would like to thank Rubik Poghossian and Hrachya Babujian for many stimulating discussions 
during their stay in Demokritos National Research Centre in Athens. 
G.S. would like to thank  Luis Alvarez-Gaume for stimulating discussions of $SL(2,Z)$ automorphic functions, the reference \cite{hejhal2} and kind hospitality at the Simon Center for Geometry.  This project received funding from the European Union's Horizon 2020 research and innovation programme under the Marie Sk\'lodowska-Curie grant agreement No 644121.

\end{document}